%% file: paper.tex
\documentclass[twocolumn]{aastex63}
\usepackage{amsmath}
\usepackage{amssymb}
\input{commands.tex}

\submitjournal{ApJ}
\shorttitle{AGN Stellar Populations}
\shortauthors{Jermyn et al.}

\begin{document}

\title{Effects of an Immortal Stellar Population in AGN Disks}

\correspondingauthor{Adam S. Jermyn}
\email{adamjermyn@gmail.com}

\author[0000-0001-5048-9973]{Adam S. Jermyn}
\affiliation{Center for Computational Astrophysics, Flatiron Institute, New York, NY 10010, USA}

\author[0000-0001-6157-6722]{Alexander J. Dittmann}
\affiliation{Department of Astronomy and Joint Space-Science Institute, University of Maryland, College Park, MD 20742-2421, USA}

\author[0000-0002-9726-0508]{B. McKernan}
\affiliation{Department of Astrophysics, American Museum of Natural History, New York, NY 10024, USA}
\affiliation{Center for Computational Astrophysics, Flatiron Institute, New York, NY 10010, USA}
\affiliation{Graduate Center, City University of New York, 365 5th Avenue, New York, NY 10016, USA}
\affiliation{Department of Science, BMCC, City University of New York, New York, NY 10007, USA}

\author{K.E.S. Ford}
\affiliation{Department of Astrophysics, American Museum of Natural History, New York, NY 10024, USA}
\affiliation{Center for Computational Astrophysics, Flatiron Institute, New York, NY 10010, USA}
\affiliation{Graduate Center, City University of New York, 365 5th Avenue, New York, NY 10016, USA}
\affiliation{Department of Science, BMCC, City University of New York, New York, NY 10007, USA}

\author{Matteo Cantiello}
\affiliation{Center for Computational Astrophysics, Flatiron Institute, New York, NY 10010, USA}
\affiliation{Department of Astrophysical Sciences, Princeton University, Princeton, NJ 08544, USA}

\begin{abstract}
Stars are likely embedded in the gas disks of Active Galactic Nuclei (AGN).
Theoretical models predict that in the inner regions of the disk these stars accrete rapidly, with fresh gas replenishing hydrogen in their cores faster than it is burned into helium, effectively stalling their evolution at hydrogen burning.
We produce order-of-magnitude estimates of the number of such stars in a fiducial AGN disk.
We find numbers of order $10^{2-4}$, confined to the inner $r_{\rm cap} \sim 3000 r_s \sim 0.03\rm pc$.
These stars can profoundly alter the chemistry of AGN disks, enriching them in helium and depleting them in hydrogen, both by order-unity amounts.
We further consider mergers between these stars and other disk objects, suggesting that star-star mergers result in rapid mass loss from the remnant to restore an equilibrium mass, while star-compact object mergers may result in exotic outcomes and even host binary black hole mergers within themselves.
Finally, we examine how these stars react as the disk dissipates towards the end of its life, and find that they may return mass to the disk fast enough to extend its lifetime by a factor of several and/or may drive powerful outflows from the disk.
Post-AGN, these stars rapidly lose mass and form a population of stellar mass black holes around $10M_{\odot}$. Due to the complex and uncertain interactions between embedded stars and the disk, their plausible ubiquity, and their order unity impact on disk structure and evolution, they must be included in realistic disk models.
\end{abstract}


\input{intro}

\input{pop}

\input{chem}

\input{outflow}

\input{merger}

\input{formation}

\input{migration}

\input{conclusion}

\acknowledgments

\input{acknowledgements}

\clearpage

\bibliography{refs}
\bibliographystyle{aasjournal}

\end{document}

%% file: commands.tex
\definecolor{purple}{rgb}{0.5,0,0.5}
\definecolor{darkgreen}{rgb}{0.1,0.6,0.1}
\definecolor{orange}{rgb}{1,0.6,0}


%% file: intro.tex
\section{Introduction}\label{sec:intro}

Active Galactic Nuclei (AGN) are formed of massive disks accreting onto supermassive black holes (SMBHs)~\citep{1969Natur.223..690L}.
These disks are thought to be unstable to fragmentation, allowing them to form stars~\citep{1980SvAL....6..357K,Goodman2003,Dittmann2020}, and they arise in the star-rich cores of galaxies, allowing them to capture stars via a variety of mechanisms~\citep{1991MNRAS.250..505S,Artymowicz1993,2020ApJ...889...94M,Fabj2020}.

Regardless of how they get there, stars embedded in AGN disks evolve in profoundly different ways~\citep{2021ApJ...910...94C,2021ApJ...916...48D}, potentially resulting in extremely massive stars~\citep{2021ApJ...910...94C,Wang_2021}, gravitational waves~\citep{McKernan14,2021ApJ...916L..17W}, and gamma-ray bursts~\citep{Perna:2020,Jermyn_2021,2021ApJ...916L..17W}.
Evolutionary models where the stars are treated as stationary inside an unchanging disk predict a state in which stars undergo no chemical/nuclear evolution.
This is because fresh gas from the disk replenishes their cores faster than hydrogen is burned~\citep{2021ApJ...910...94C,2021ApJ...916...48D}.
We call such stars ``immortal''.

Here we provide rough, order-of-magnitude estimates of the number of immortal stars embedded in the inner regions of AGN disks ($r \la 0.1\,\rm pc$, Section~\ref{sec:pop}).
In doing so we only consider stellar captures, as star-formation in AGN disks remains poorly understood.
We emphasize that these estimates are highly uncertain, but nevertheless allow us to make several interesting predictions.
For instance, embedded immortal stars may be numerous enough to significantly enrich the disk with helium, and we estimate the significance of this effect in Section~\ref{sec:chem}.
In Section~\ref{sec:outflow} we examine how these stars evolve as the disk dissipates and their reservoir of fresh gas depletes, finding that they may either extend the life of the disk or else drive massive outflows from the disk.
In the former case we predict that up to half of AGN disks may be presently fed by mass loss from massive stars, releasing mass which these disks lost to stars in an earlier era.
In Section~\ref{sec:merger} we estimate the merger rates of these stars, and speculate on the outcomes of such mergers.
In Section~\ref{sec:sfr} we naively estimate the star formation rate, and show that \emph{in-situ} formed stars could well dominate over the captured population, which points to the need for self-consistent disk models which account for star formation.
We then conclude in Section~\ref{sec:conclusions}.

%% file: pop.tex
\section{Stellar Population}\label{sec:pop}

\begin{figure*}
\centering
\includegraphics[width=0.95\textwidth]{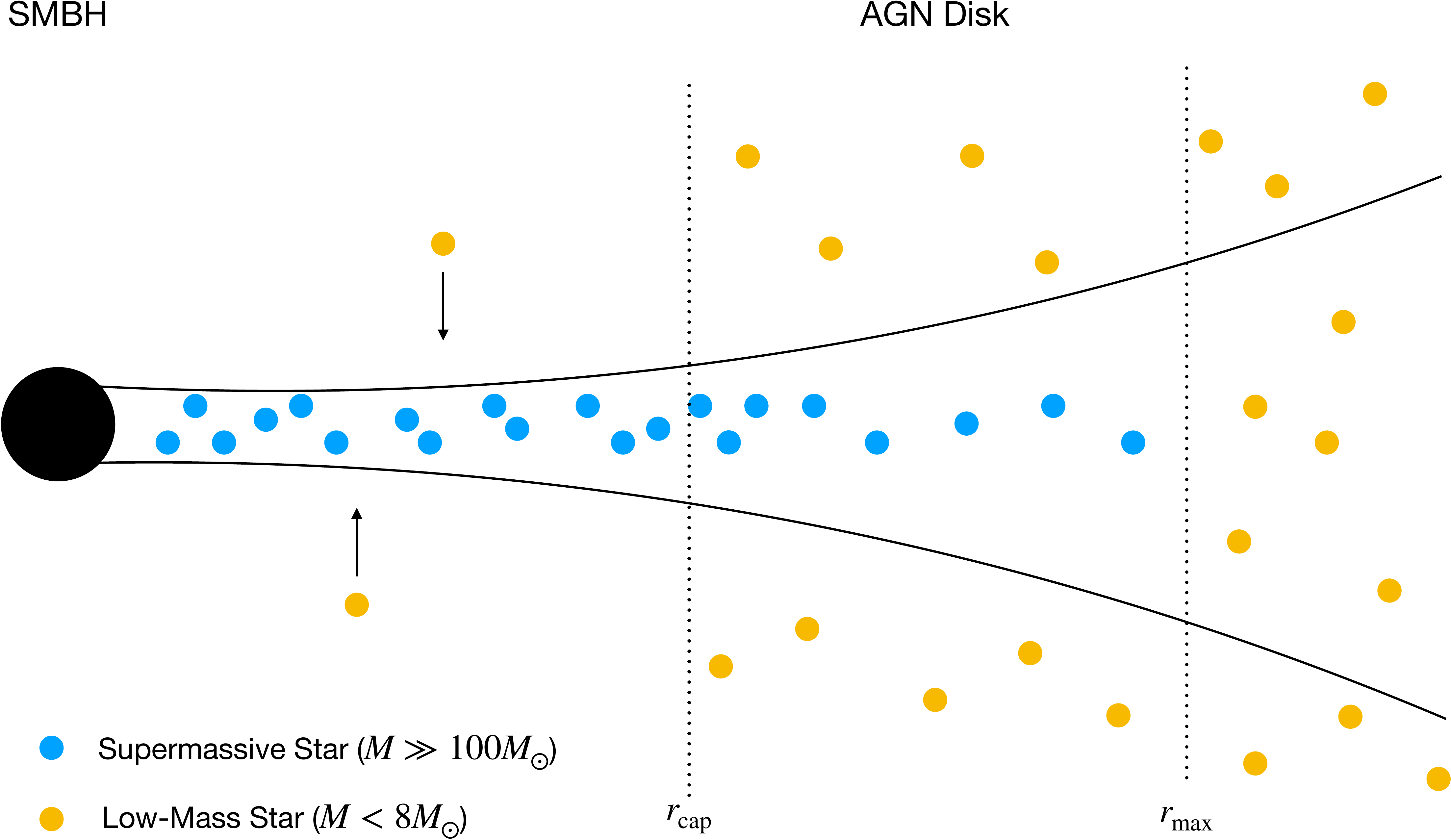}
\caption{A schematic of an AGN disk with an embedded stellar population is shown. Inside the capture radius $r_{\rm cap}$ most low-mass stars are captured by the disk, and outside the capture radius few stars are captured. Inside the radius $r_{\rm max}$ stars that are captured rapidly grow to be high-mass immortal stars, and outside it they accrete relatively little and pass through the disk unperturbed.}
\label{fig:schema2}
\end{figure*}

\subsection{Stellar Models}

The evolution of stars embedded in AGN disks is characterized by accretion of disk material at the Bondi-Hoyle-Littleton rate~\citep{2021ApJ...910...94C,2021ApJ...916...48D}.
Following~\citet{2021ApJ...910...94C}, immortal stars live in regions with
\begin{align}
	\rho > 10^{-17}\mathrm{g\,cm^{-3}}\left(\frac{c_s}{3\mathrm{km\,s^{-1}}}\right)^{3}\left(\frac{t_{\rm disk}}{\mathrm{Myr}}\right)^{-1},
	\label{eq:crit2}
\end{align}
such that a $1 M_\odot$ star can more than double in mass in the disk lifetime $t_{\rm disk}$, which is much shorter than the chemical evolution time for such a star.
Here $\rho$ is the gas density in the disk and $c_s$ is the sound speed.
In Section~\ref{sec:disk} we discuss the likely range of disk lifetimes, which is centered on $t_{\rm disk} \sim 1\,\mathrm{Myr}$.

Near the SMBH tidal effects limit accretion, so we additionally require that the density be great enough to overcome this reduction, such that 
\begin{align}
\frac{\rho}{\mathrm{g\,cm^{-3}}} > 6\times 10^{-3} \left(\frac{c_s}{3\mathrm{km\,s^{-1}}}\right)^{-1} \left(\frac{\Omega}{\mathrm{s^{-1}}}\right)^{4/3},
\label{eq:crit}
\end{align}
where
\begin{align}
\Omega=\sqrt{\frac{G M_\bullet}{r^3}}
\end{align}
is the Keplerian angular velocity, $M_\bullet$ is the mass of the central black hole, $G$ is the gravitational constant, and $r$ is the orbital radius~\citep[][Eq. 22]{2021ApJ...916...48D}. The difference between equations (\ref{eq:crit2}) and (\ref{eq:crit}) stems from the accretion rate onto stars scaling as $\dot{M}\propto\rho c_s R_H^2\propto\rho c_s \Omega^{-4/3}$ rather than $\dot{M}\propto\rho c_s R_B^2\propto \rho c_s^{-3}$ when the Hill radius $(R_H=r(GM_*/3M_\bullet)^{1/3})$ \citep{hill1878} is smaller than the Bondi radius $(R_B=2GM_*c_s^{-2})$ \citep{1952MNRAS.112..195B}.
Here $M_*$ is the stellar mass.

Using the definition of the Schwarzschild radius,
\begin{align}
	r_s = \frac{2 G M_\bullet}{c^2} \approx 10^{-5}{\rm pc} \left(\frac{M_\bullet}{10^8M_\odot}\right),
\end{align}
we may write equation~\eqref{eq:crit} instead as
\begin{align}
\frac{\rho}{\mathrm{g\,cm^{-3}}} > 6\times 10^{-7} \left(\frac{c_s}{3\mathrm{km\,s^{-1}}}\right)^{-1} \left(\frac{r}{r_s}\right)^{-2}\left(\frac{M_\bullet}{10^8 M_\odot}\right)^{-4/3},
\end{align}
once more subject to criterion~\eqref{eq:crit2}.

\subsection{Disk Model}\label{sec:disk}

We employ the fiducial disk model from figure~2 of~\citet{2003MNRAS.341..501S} (SG), with an SMBH mass of $10^8M_\odot$ and a gas accretion rate of order $1 M_\odot/\mathrm{yr}$.
This model has multiple solutions for the temperature profile for $r \ga 3\times 10^5 r_s$.
We only make use of the density and scale-height, however, and both are single-valued over the whole domain, so for our purposes this model suffices and is unambiguous.

In the SG disk model the density outside of $\sim 10^3 r_s$ scales as $\rho \approx 10^{-9}\mathrm{g\,cm^{-3}} (r/10^3 r_s)^{-3}$, so we can solve for the radius at which stars cease to be immortal (Figure~\ref{fig:schema2}) and find
\begin{align}
r_{\rm max} \approx 2\times 10^6 r_s.
\end{align}
from equation~\eqref{eq:crit} and
\begin{align}
r_{\rm max} \approx 5\times 10^5 r_s \left(\frac{t_{\rm disk}}{\mathrm{Myr}}\right)^{1/3}.
\end{align}
from equation~\eqref{eq:crit2}.
The relevant radius is the smaller of these, which is typically the latter for this disk model.

Because we are only interested in immortal stars we restrict our study to regions with $r < r_{\rm max}$.

\subsection{Stellar Capture}

Roughly $10^{6} M_\odot$ stars live in the central parsec of the Milky Way~\citep{2010ApJ...718..739M}.
Similar estimates are found for other nuclear star clusters (NSCs)~\citep{B_ker_2008}.

A Salpeter IMF gives a mean stellar mass of $\approx 0.7 M_\odot$~\citep{1955ApJ...121..161S}, implying that there should be $\sim 1.4 \times 10^6$ stars in this region.
We round this down to $10^6$ stars of interest because the lowest-mass stars will never accrete enough to become immortal.

The distribution of stars is likewise uncertain, but observations of the Milky Way suggest a plausible model is a constant surface density in the inner parsec~\citep{2010ApJ...718..739M}, so we use a constant $\Sigma \approx 3\times 10^5 \mathrm{pc}^{-2}$ going forward.

\citet{Fabj2020} showed that with the fiducial SG disk model most stars of solar-type and earlier are captured by the disk in $10\rm Myr$ for radii less than $5\times 10^3 r_s$.
For shorter lifetimes of $0.1\,\rm Myr$ stars are instead captured out to $r \la 10^3 r_s$.
These lifetimes were chosen to span the range from some of the shortest estimates that have been proposed~\citep{10.1093/mnras/stv1136,10.1093/mnrasl/slv098} up to $\sim 1\%$ of the longest lifetimes consistent with energetic bounds~\citep{1982MNRAS.200..115S,2004MNRAS.351..169M}.
We will generally refer to the \emph{capture radius} as the radius out to which a typical solar-type star is captured (Figure~\ref{fig:schema2}), and use
\begin{align}
1000 r_s \la r_{\rm cap} \la 5000 r_s.
\label{eq:rcap}
\end{align}
Note that this capture radius is smaller than $r_{\rm max}$, so we expect all captured stars to be immortal, subject to concerns about the gas supply which we discuss in the next section.

Neglecting the dependence of capture time-scales on the SMBH mass, we estimate that a short-lived AGN disk captures
\begin{align}
N_{\rm short} \sim \pi (10^3 r_s)^2 \Sigma \approx 10^3 \left(\frac{M_\bullet}{10^8 M_\odot}\right)^2
\end{align}
stars, and a long-lived one captures
\begin{align}
N_{\rm long} \sim \pi (5\times 10^3 r_s)^2 \Sigma \approx 2\times 10^4 \left(\frac{M_\bullet}{10^8 M_\odot}\right)^2,
\end{align}
where $M_\bullet$ enters only by setting the scale $r_s$.

\subsection{Population Estimates}

\subsubsection{Short-lived disks ($\tau \sim 0.1\,\rm Myr$)}

Even for very short-lived disks, captured stars have a chance to accrete and become massive because the gas density is so high ($\rho \ga 10^{-9}\,\mathrm{g\,cm^{-3}}$ inside $r \la 10^3 r_s$).
In these regions and over that time-span roughly $N_{\rm short} \sim 10^3$ are captured and grow to $M \sim 300 M_\odot$~\citep{2021ApJ...910...94C}.

For all of these stars to become immortal requires $\sim 3 M_\odot\,\mathrm{yr}^{-1}$ across the disk, which is three times the total $\dot{M} \sim 1 M_\odot\,\mathrm{yr}^{-1}$ of the disk, so in practice the disk structure will be substantially altered by the presence of embedded stars~\citep{2021arXiv210707519G}, and we expect the immortal population to be limited by the available gas in these short-lived disks.
Most of this gas will be consumed by the innermost stars because these grow the fastest, so we expect $\sim 300$ immortal stars and $\sim 700$ mortal stars with slightly enhanced masses.
These mortal stars, unlike the more massive immortal ones, will live well past the dissipation of the disk and so could explain the observed population of $\sim 10^2$ massive stars in the center of the Milky Way~\citep{1995ApJ...447L..95K,2003ApJ...590L..33L,2006ApJ...643.1011P}, where an AGN phase is thought to have happened about $2-8\,\mathrm{Myr}$ ago~\citep{2010ApJ...724.1044S,2019ApJ...886...45B}.

\subsubsection{Long-lived disks ($\tau \sim 10\,\rm Myr$)}

For lifetimes of at least a few million years a population of order $N_{\rm long} \sim 2\times 10^4$ stars are captured.
As before these stars accrete and become massive.
Once more this process could consume more than the total disk $\dot{M}$, but this time there is just about enough $\dot{M}$ of gas in the disk to accommodate this growth: these stars require of order $0.6 M_\odot\,\mathrm{yr}^{-1}$ of gas to reach $300 M_\odot$, which is a bit less than the disk total of $1 M_\odot\,\mathrm{yr}^{-1}$.
Hence we expect the full population of $\sim 2\times 10^4$ captured stars to be immortal.

%% file: chem.tex
\section{Chemical Evolution}\label{sec:chem}

Immortal stars never evolve past hydrogen burning, so they primarily process hydrogen into helium.
Because these stars are extremely massive they have large convective cores, which mix the resulting helium much of the way to the surface.
In the radiative regions outside the core we expect that instabilities caused by radiation pressure, rotation~\citep{2021ApJ...914..105J}, and mass loss will result in further mixing, so that the whole star ends up well-mixed and the surface helium abundance matches that of the core~\citep[see also section 4.4 of][]{2021ApJ...910...94C}.

The net result is that the super-Eddington winds which carry mass away from these stars should be enhanced in helium relative to accreting material.
This enriches the disk in nucleosynthetic products, principally helium, and our aim here is to estimate the magnitude of this effect.

Because they exist via a balance between super-Eddington mass loss and accretion, the luminosities of these stars are tuned to near the Eddington limit
\begin{align}
	L = \Gamma L_{\rm Edd} \approx 3\times 10^4 L_\odot \left(\frac{M}{M_\odot}\right),
\end{align}
where $\Gamma \ga 1$ is the Eddington ratio, which we expect to be of order but somewhat greater than unity.
For now we fix $\Gamma=1$, but recall that if it is $2-3$ then the nuclear burning rate is $2-3$ times larger.

\citet{2021ApJ...910...94C} see stars reaching masses of $5\times 10^2-2\times10^3M_\odot$, though effects considered by~\citet{2021ApJ...916...48D} likely lower this somewhat, so using $300 M_\odot$ we find $L \approx 10^7 L_\odot$.
Most of the energy in hydrogen burning is emitted in the form of gamma rays and positrons rather than neutrinos, so most of the rest mass difference between $4^{1}\rm H$ and $1^{4}\rm He$ is converted into heat, resulting in $27\rm MeV$ per $^{4}\rm He$ nucleus formed~\citep{1958ApJ...127..551F}.
It follows that immortal stars produce $^{4}\rm He$ at a rate of
\begin{align}\label{eq:helium}
	\dot{M}_{\rm He} \approx \frac{L}{27\mathrm{MeV}/4 m_p} \approx 10^{-4}M_\odot\,\mathrm{yr}^{-1},
\end{align}
where $m_p$ is the proton mass.
Immortal stars return all of the helium they produce to the disk, so each star contributes helium to the disk at a rate of $\dot{M}_{\rm He}$.

For short-lived disks this produces a flux of $\mathrm{H}\rightarrow\mathrm{He}$ of order $\dot{M}_{\rm He} N_\star \approx 3\times 10^{-2} f_\star M_\odot\,\mathrm{yr}^{-1}$, which is $3\%$ of the mass flux of gas through the disk.
Short-lived disks could then have helium mass fractions near the SMBH as high as $25\%$.

The same calculation for long-lived disks gives an even more extreme result of $2 M_\odot\,\mathrm{yr}^{-1}$, which exceeds the disk $\dot{M}$.
Naively then, long-lived disks could have mass fractions approaching unity, at which point other physics becomes important, as we discuss below.

Note that these enhancements scale with the equilibrium mass of immortal stars.
If these stars only reach $100 M_\odot$ rather than $300 M_\odot$ the enhancements in mass fractions we expect are $1\%$ and $60\%$ respectively.

In practice even for more massive stars the helium enhancements may be somewhat smaller.
As the hydrogen mass fraction in the disk falls, immortal stars become more helium-rich, and so their nuclear burning rates rise.
This makes it harder to replenish hydrogen to their cores quickly enough to prevent them from evolving.

We do not know the result of this feedback, but one possibility that seems at least plausible is that the immortal stars end up restricted to a narrow annulus in the disk.
Primordial-abundance material enters the band and is processed, resulting in a helium abundance gradient in the disk.
Inward, past a certain point, immortal stars may not be able to exist.
Stars in the inner regions could evolve rapidly, gaining and then shedding mass, exploding, and leaving compact objects behind.
This would prevent the central regions from becoming totally helium-dominated, would reduce the overall stellar population, and would allow for more compact remnants to form as stars migrate out of the immortal band.

Regardless, significant helium enhancement seems likely in AGN disks.
Because the enrichment we predict is so large, and because the enriched matter flows in towards the SMBH, it should be possible to test our model with observations of helium abundances in the broad line region.
Should observations contradict this prediction, then we have likely either overestimated the stellar population, neglected an important channel whereby stars can be destroyed/ejected, or else neglected some factor which precludes immortal stars from existing in the first place.

Finally, while we have focused on helium enrichment here, there is also evidence of metal enrichment in AGN disks~\citep{1999ARA&A..37..487H,2006A&A...447..157N,2021ApJ...910..115S}.
In particular, it appears that AGN disks have nearly constant metallicity over cosmic time~\citep{2018MNRAS.480..345X}, suggesting an intrinsic source of this enrichment which could point to embedded stars.
Stellar models suggest relatively modest metal yields in the immortal phase, of order $10^{-4}$ times that of helium~\citep{2021ApJ...916...48D}, indicating that immortals are unlikely to contribute much to the metal fraction.
Interestingly, we expect significantly higher~\citep[$\sim 0.03 M_\odot\,\mathrm{yr}^{-1}$ per star,][]{2021arXiv211206151T} metal yields for \emph{mortal} stars embedded in the disk because these still become massive but are still able to evolve towards later stages of nuclear burning in addition to further metal enrichment from supernovae~\citep[see also][]{Artymowicz1993}.
It could then be that mortal stars further out in the disk contribute the most to metal-enrichment, while helium production is the domain of the immortals.

%% file: outflow.tex
\section{Stellar Winds}\label{sec:outflow}

\subsection{AGN Feedback}

\begin{figure*}
\centering
\includegraphics[width=\textwidth]{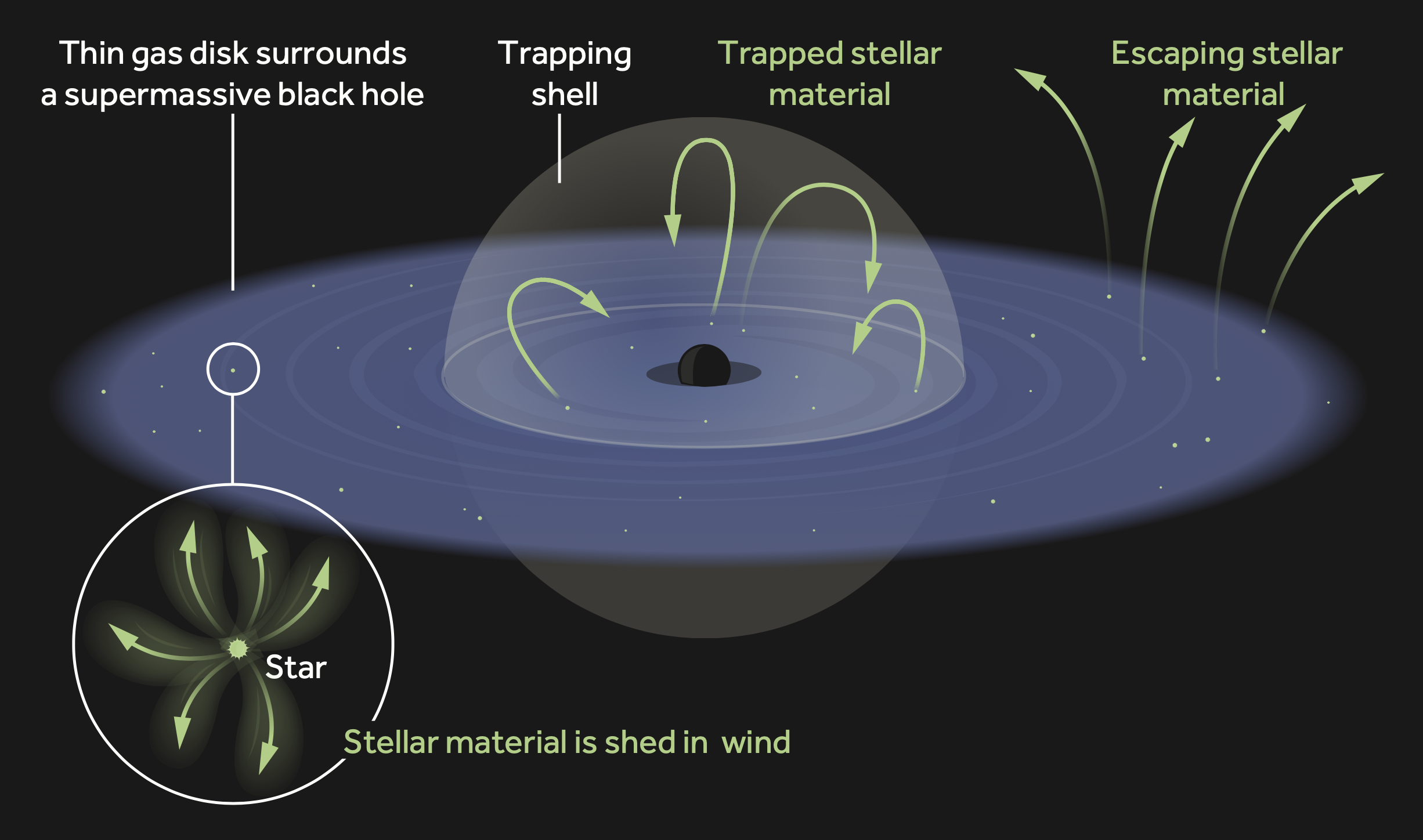}
\caption{Our wind-fed disk scenario is shown schematically. A thin gas disk surrounds an SMBH. Embedded stars shed material in a wind. When these are further out, where the escape velocity from the SMBH is low, the wind forms an outflow from the disk. When they are closer in the stellar winds cannot escape, and serve to return material to the disk. When the disk ceases to be fed from the outside it gradually spirals inward and truncates to the trapping shell, where the escape velocity from the disk equals the wind speed. The disk can live for millions of years in this wind-fed phase.}
\label{fig:schema}
\end{figure*}

Feedback from AGN winds is likely important for sculpting the inner regions and bulges of host galaxies in the $\Lambda$CDM model of structure formation \citep[e.g.][]{Silk98,King03}, and the mass and velocity in AGN outflows is generally believed to be associated with a conservation of angular momentum from accretion of mass onto the SMBH \citep[see e.g.][for a recent review]{Harrison18}.

Even before the disk dissipates, embedded immortal stars could play an important role in supplying material for feedback.
Because their accretion and mass-loss rates balance, and are of order the disk $\dot{M}$, they eject of order $\sim 1 M_\odot\,\mathrm{yr}^{-1}$ in stellar winds.
The winds from these stars are expected to reach typical velocities set by the escape velocity from the star~\citep{2008MNRAS.389.1353V}.
Using a typical radius of $R_\star \sim 10 R_\odot$~\citep{2021ApJ...910...94C}, we find speeds of $\sim 3\times 10^3\mathrm{km\,s^{-1}}$, comparable to the local Keplerian velocity.
In the outer parts of the nucleus, where the escape velocity from the SMBH is smallest, stellar winds would drive outflows from the disk and thereby add to AGN feedback.
This is shown schematically in Figure~\ref{fig:schema}.

At smaller disk radii, where SMBH escape speeds are larger, the same outflows may simply resupply the disk and extend its lifetime, which we discuss in more detail below. If a significant fraction of the mass of stellar outflows escapes the nucleus, however, the lifetime of the disk would \emph{not} be extended---so detailed modeling of embedded objects is critical to our understanding of both feedback and lifetimes of AGN disks.

We call the transition radius, where stellar winds are marginally bound, the trapping radius
\begin{align}
	r_{\rm trap} \approx \frac{M_\bullet}{M_\star}R_{\star} \sim 10^7 R_\odot \sim 2\times 10^4 r_s,
\end{align}
where we have used a typical radius of $R_\star \sim 10 R_\odot$~\citep{2021ApJ...910...94C}.
Note that this estimate implies that
\begin{align}
	r_{\rm cap} \ll r_{\rm trap} \ll r_{\rm max},
\end{align}
though this statement is dependent on the disk model and capture physics and so may not hold in general.

These outflows are also interesting for supplying a volume filling gas above and below the disk, which then acts as a source of drag on orbiters passing through them. It may therefore drive disk capture of NSC objects.
Along similar lines, if AGN disks originate from discrete `pulses' of low angular momentum gas which falls out of the torus or wider gas reservoir, there is no required connection between the inner disk and the torus. Winds from transitioning immortal stars could be responsible for filling the gap between these two regions.

\subsection{Disk Lifetimes}

Immortal stars are characterized by a balance between mass accretion and stellar winds, holding their total mass in equilibrium~\citep{2021ApJ...910...94C}.
Towards the end of the life of an AGN disk the gas density presumably falls, reducing the accretion rate.
We expect then that the region in which immortal stars can exist moves inwards.
If this occurs faster than stars migrate then individual stars can cross the threshold and become mortal.

Without a rapid supply of fresh fuel, these newly-mortal stars begin to evolve.
Their cores deplete in hydrogen and their specific nuclear burning rates rise, resulting in much more rapid mass loss.
The result is that the vast majority of the mass in these stars, at least $90\%$, is shed on time-scales of $\sim 0.3~\rm Myr$~\citep{2021ApJ...910...94C}.
We found in Section~\ref{sec:pop} that the stellar population likely contains a large fraction of all the mass in the disk, having used up a substantial fraction of the gas mass flux.
If the stellar winds are slow enough that this gas remains in the disk, the mass being returned to the disk can potentially extend its lifetime by an amount of order unity.
This is shown inside the ``trapping shell'' in Figure~\ref{fig:schema}.

For instance, suppose the disk is short-lived, such that the infalling gas from the outer reaches of the disk ceases after $\tau \sim 0.1\rm Myr$.
This is fast relative to any disk migration time-scale (Section~\ref{sec:migration}) and the mass-shedding time, so all of the immortal stars in the disk will start to return mass to the disk.
During the lifetime of the disk stars accreted of order its $\dot{M}$, so they contain of order the gas mass of the disk.
They can return gas to the disk at a rate $\dot{M}_{\rm stars} \sim M_{\rm stars} / 0.3~\rm Myr \sim 0.5 \dot{M}_{\rm disk}$, and could well maintain the disk at a reduced gas density.
This would continue for roughly $0.3\,\rm Myr$, at which point the stars run out of mass, and so extend the lifetime of the disk by a factor of four.

On the other hand if the disk is long-lived ($\tau \sim 10\,\rm Myr$) then as it dissipates stars will become mortal in sequence, from the outside of the disk in.
Because stars shed their mass faster than the disk dissipates they can supply enough mass to maintain the original disk $\dot{M}$, and indeed the system likely regulates towards this point.
If the stars fail to restore the disk $\dot{M}$ then the disk density falls faster, even more stars start losing mass, and so the star $\dot{M}$ rises.
By contrast if the stars return too much mass to the disk then the disk density rises, causing some of them to become immortal once more, reducing the stellar $\dot{M}$.
Thus we expect the stars to act to hold the disk density profile and gas accretion rate constant.
This can proceed until the system runs out of immortal stars.
If half of the mass of the system is in such stars, as seems likely given that one third of all the gas accreted by the disk ends up in stars, then this can double the lifetime of the disk.
Greater extensions are possible, but only if the mass of the disk is overwhelmingly concentrated in stars, in which case other effects including stellar interactions and dynamics become important.

These star-wind-fed AGN disks could appear quite different from younger, externally-fed disks.
The gas density would be lower, there would be no external torques feeding in mass from the outer edge and so the angular momentum balance could be quite different, and the disk could potentially have a surface density distribution that tracks the radial stellar density.

A zeroth order picture of such a disk might consist of a comparatively compact, standard thin disk extending out to the edge of where immortal stars can live ($r_{\rm max} \sim 2\times 10^6 r_s$).
Above and below this disk would be a volume-filling flow with a spiral velocity structure, tracing the orbits of the stars that feed it.
This flow would have velocities below the SMBH escape velocity and so would not outflow, but would gradually settle back into the midplane and replenish the disk.
Such a structure could mimic a close-in torus observationally and could act as a source of broad emission lines.
These are just some leading order considerations, however, and our understanding of such wind-fed disks would benefit from further study.

\subsection{Post-AGN Evolution}

By the time the disk has dissipated all immortal stars are mortal.
Massive outflows coupled with stellar evolution drive these remaining massive stars to produce stellar mass black holes (BH) around $10M_{\odot}$ \citep{2021ApJ...910...94C}, though with some variation with the uncertain physics of mass loss.
There is a hint of a mass excess around $10M_{\odot}$ in the underlying black hole mass function inferred by LIGO-Virgo from O3a \citep{O3a,2021arXiv211103634T}.
Such an excess could be the signature of the population of BH produced by immortals at the end of the AGN phase, and would imply that AGN play an important role in generating the mergers observed by LIGO-Virgo.

%% file: merger.tex
\section{Stellar Mergers}\label{sec:merger}

The differential migration\footnote{See the discussion of migration rates in Section~\ref{sec:migration}. Importantly, objects of different mass migrate through the disk at different rates.} of black holes, other stellar remnants and stars in AGN disks may lead to accelerated binary formation and possibly mergers driven by gas or dynamical hardening \citep[e.g.][]{McKernan2012}. The resulting compact-object mergers can likely be detected with LIGO-Virgo in gravitational waves \citep[e.g.][]{McKernan14,Bartos17,Stone17,Secunda20,Tagawa20,McKernan20} and possibly in electromagnetic radiation \citep[e.g.][]{McKernan19,Graham20,Palmese21,Perna21}. 

The same basic processes of differential migration, binary formation and merger also apply to immortal stars in AGN disks. Importantly, as immortal stars migrate through the AGN disk they will encounter other immortal stars, black holes and other remnants.
Immortal stars passing close enough to another massive object, within a mutual Hill sphere\footnote{i.e. so that binding energy is greater than the relative kinetic energies} will form a binary. The resulting binary can then be hardened or softened by gas torques or dynamics, or both, depending on the distribution of gas around the binary and the rate and type of tertiary encounters \citep[e.g.][]{Baruteau11,Leigh18,YapingLi21}. The details of what may happen in the case of an encounter between an immortal star and another massive object require detailed 3D hydrodynamic modeling including radiative feedback, so here we just discuss some possibilities and considerations impacting the different types of merger.

\subsection{Immortal star $+$ Immortal star}

Suppose there are $N_{\star}$ immortal stars scattered across the inner AGN disk, all migrating with a typical time-scale $t_{\rm m}$ which exhibits some power-law dependence on mass.
Then objects of different masses will encounter each other at a rate up to ${\cal{R}_{\rm IM-IM}}\sim N_{\star}/t_{\rm m}$.
If there are places in the AGN disk where migration slows and pile-up can occur \citep[e.g. migration traps][]{Bellovary16} then repeated mergers involving most of the immortal stars can occur during pile-up at the trap \citep{Secunda20}.
Given a large sample of $N_{\rm AGN}$, then the rate of such mergers in the sample would be $N_{\rm AGN}{\cal{R}_{\rm IM-IM}}$. Thus, if detectable, the rate of immortal star mergers in large-scale surveys including e.g. ZTF \citep{2021ApJ...920...56F} could be $\sim \mathcal{O}(10^2)\mathrm{yr}^{-1}(N_{\rm AGN}/10^{6})(\overline{N}_{\star}/10^{2})(\overline{t}_{\rm m}/1{\rm Myr})$, where $\overline{t}_{\rm m},\overline{N}_{\star}$ are the average migration times and numbers of immortal stars per AGN disk respectively.

Because immortal stars are characterized by a balance between accretion and super-Eddington winds, there is a characteristic equilibrium mass.
As a result, the merged object must quickly shed roughly half of its mass to return to this equilibrium.
If this does not occur dynamically during the merger itself, we expect it to occur on a time-scale
\begin{align}
	t_{\rm s} &= \frac{M_\star v_{\rm esc}^2}{L_{\rm Edd}} = \frac{M_\star \kappa}{8\pi c R_\star} \nonumber \\
	&\approx 10^4\mathrm{yr}\left(\frac{M_\star}{300M_\odot}\right)
	\left(\frac{R_\star}{10R_\odot}\right)^{-1}\left(\frac{\kappa}{0.4\mathrm{cm^2\,g^{-1}}}\right).
\end{align}
So in this sense immortal stars which merge do not accumulate mass.
Rather two stars become one, and the mass of the other is returned to the disk.

The high merger rate, and unstable nature of the merger remnants, suggests that detailed hydrodynamic studies of immortal star mergers in AGN disks are necessary for understanding the number and type of post-AGN stars and stellar remnants.

\subsection{Immortal star $+$ Compact Object} 

Black holes (BH) embedded in the AGN disk could merge with immortal stars.
The result of this process is highly uncertain, but it could result in a quasi-star, in which a black hole is embedded in a star and radiation pressure from accretion onto the black hole serves to prop up the stellar envelope~\citep{Begelman08}.
Similarly, embedded neutron stars could merge with immortal stars, forming Thorne-$\dot{\rm Z}$ytkow objects, Red Supergiants with neutron stars for cores~\citep{1977ApJ...212..832T}.
Both such exotic remnants would be extremely luminous, and could play an important role in shaping the AGN disk, though existing models of their structure and evolution remain quite uncertain.
One intriguing possibility is that immortal stars could successively capture two BHs, which would form a binary that could merge within the immortal star.
This is similar to the scenario of successive mergers of BH as modeled in e.g.~\citet{Secunda20,McKernan20}.
The remnant of such a BBH merger would only escape the immortal star if a sufficiently large kick occurs at merger.
This possibility warrants further attention as it could significantly impact the parameters expected from the BBH detected by LIGO-Virgo from the AGN channel. 

\subsection{Immortal star $+$ SMBH}

If there are no migration traps then immortal stars could end up very near the SMBH.
Gradual migration towards the SMBH is insufficient to tidally disrupt the star, rather an immortal could experience Roche-lobe overflow onto the SMBH in the innermost disk \citep[e.g][]{2013MNRAS.434.2948D,2017ApJ...844...75M,2021arXiv210713015M}.
Such a state might yield quasi-periodic oscillations and even an enhanced accretion episode onto the SMBH, yielding a temporary boost in $\dot{M}_{\rm SMBH}$, accompanied by a color change as the inner disk temperature ($\propto \dot{M}_{\rm SMBH}^{1/4}$) and luminosity ($\propto \dot{M}_{\rm SMBH}$) increase. The rate of such events could be as much as $\sim \mathcal{O}(10^{2}){\rm yr^{-1}} (N_{\rm AGN}/10^{6})(\overline{N}_{\star}/10^{2})(\overline{t}_{\rm m}/1{\rm Myr})$. Such state-changes in AGN accretion rate and quasi-periodicity should be detectable in large optical/UV surveys \citep[e.g.][]{Graham17}.

Dynamical encounters between immortal stars and other migrators, particularly other immortal stars in binary-single encounters, can lead to chaotic encounters and interactions including large-angle three-body scatterings~\citep{2021arXiv211003698W}. There is an AGN loss-cone which results from the scatterings of stars and other objects onto the SMBH \citep[e.g.][]{Starfall}. If an immortal star is scattered onto, or close enough to, the SMBH, a tidal disruption event (TDE) can occur. Unlike the main sequence stars believed to be involved in most TDEs, an immortal star has a large enough radius and is massive enough that it can be tidally disrupted by SMBH up to $\sim 10^{9}M_{\odot}$. However, immortal stars may be massive enough that they can only be scattered through large angles by low mass IMBH in the disk. The resulting TDE is super-luminous, since none of the stellar ejecta escapes; rather it collides with AGN disk gas at small disk radii and drives a high accretion rate event.
These events should be more luminous than TDEs in quiescent galactic nuclei and have shorter characteristic timescales \citep{Starfall}.

A tidal disruption of an Immortal star should yield a spectacular flare followed by a change in the equilibrium state of the disk lasting many years. This is an amplified version of the flares expected in regular TDEs in AGN \citep{Starfall}. The detection of such super-TDE flaring around large mass SMBH (up to $\sim 10^{9}M_{\odot}$) followed by a change in the continuum emission level would represent a clear detection of the tidal disruption of a super-massive star and subsequent disruption of the inner AGN disk. The absence of such flaring in a large sample of AGN over an extended period of time would put constraints on the rate of occurrence and scattering of Immortal stars in AGN disks. We are presently searching for evidence of such TDE-like events in the O($10^{6}$) AGN monitored by ZTF.

%% file: formation.tex
\section{Star Formation}\label{sec:sfr}

Star formation rates in AGN disks are highly uncertain.
Following~\citet{2004ApJ...608..108G}, we assume that star formation happens outside $r_{\rm min,\,SF} = r_{sg} \sim 2000 r_s$ via fragmentation.

The fragmentation instability is axisymmetric, so assuming one star forms per instability wavelength in the radial direction per Kelvin-Helmholtz time, we find
\begin{align}
	\frac{d\dot{N}}{d\ln r} \approx 3\times 10^{-4}\frac{r}{h}\mathrm{yr}^{-1},
\end{align}	
where we have used $3300\,\mathrm{yr}$ for the Kelvin-Helmholtz time~\citep{1984ApJ...280..825B,2004ApJ...608..108G} and $\lambda \approx 1/h$ for the instability wavelength~\citep{2004ApJ...608..108G}.

For our fiducial disk model, the region of interest ($r_{\rm min\,SF} < r < r_{\rm max}$), has $h/r$ between $10^{-2}$ and $0.3$, so we take as a typical value of $r/h \approx 30$ and parameterize our remaining uncertainty as
\begin{align}
	\frac{d\dot{N}}{d\ln r} \approx 10^{-2}f_\star \mathrm{yr}^{-1},
	\label{eq:dNdlnR}
\end{align}	
where $f_\star$ is an order-unity parameter accounting for the crudeness of our model.
Integrating over the range from $r_{\rm min}$ to $r_{\rm max}$ we find
\begin{align}
	\dot{N} \approx 7\times 10^{-2}f_\star \mathrm{yr}^{-1}.
\end{align}	

For any plausible disk lifetime $\tau \ga 10^5\,\mathrm{yr}$ the number of stars formed in the disk is greater than the number captured.
We thus expect the population of AGN stars to be dominated by \emph{in-situ} formation rather than stellar capture.
We caution however that this is a very simplistic estimate, which we include only to highlight the need for more work on self-consistent disk models incorporating star formation, along the lines of~\citet{2003astro.ph..7084L,2021arXiv210707519G}.

Stars can also form beyond $r_{\rm max}$.
Such star are not immortal, but rather form, accrete material, chemically evolve, and die, likely with lifetimes comparable to the typical $10-100\mathrm{Myr}$ for massive stars.
These stars likely form at similar rates to the immortal ones, as the scalings in equations~\eqref{eq:dNdlnR} are similar in the regions just beyond $r_{\rm max}$.
Because these stars live so long they will endure past the dissipation of the disk and so could potentially explain observations of a top-heavy IMF in the center of the Milky Way~\citep{2010ApJ...708..834B}.
By the same token, because these stars do not grow extreme masses they have a much smaller influence on the disk than the inner immortal population, and we expect them to contribute little to e.g. chemical enrichment of the disk.

%% file: migration.tex
\section{Migration Rates}\label{sec:migration}

Because migration is highly uncertain we have not accounted for it in our population estimates, but we provide an estimate of the migration rate here to demonstrate that, depending on the details of the disk structure, the effects of migration range from irrelevant to directly setting the number of stars in the disk.

We begin by estimating the rate at which stars migrate through the disk due to linear corotation and Lindblad torques, giving the migration time-scale \citep[e.g.][]{1979ApJ...233..857G,1997Icar..126..261W,2002ApJ...565.1257T}
\begin{align}
t_{\rm m}  \approx \frac{r^2 \Omega M_\star}{2 \Gamma_0} \approx \Omega^{-1}\frac{M_\bullet}{r^2 \Sigma}\left(\frac{M_\star}{M_\bullet}\right)^{-1}\left(\frac{h}{r}\right)^{2},
\end{align}
where  $h$ is the disk pressure scale-height, $\Gamma_0 = (M_\star/M_\bullet)^2(h/r)^{-2} \Sigma r^4 \Omega^2$, and $\Sigma$ is the disk surface density.
Using $Q \sim \Omega c_s / G \Sigma\sim 1$ we find
\begin{align}
t_{\rm m}  &\approx \frac{G M_\bullet}{\Omega^2 r^2 c_s}\left(\frac{M_\star}{M_\bullet}\right)^{-1}\left(\frac{h}{r}\right)^{2}\approx \Omega^{-1}\left(\frac{M_\star}{M_\bullet}\right)^{-1}\left(\frac{h}{r}\right).
\label{eq:tm}
\end{align}
For $h/r \sim 10^{-2}$, $M_\star \sim 3\times 10^{-6} M_\bullet$, and $r\sim r_{\rm cap} \sim 0.03{\rm pc}$ we find a timescale of $3\times 10^4\,\rm yr$, so migration is likely quite important, and could substantially reduce the stellar population in the disk.

Migration can be slowed significantly if gaps open in the disk, decreasing the local surface density when the parameter $K = (M_\star/M_\bullet)^2 (h/r)^{-5} \alpha^{-1} \gtrsim K_t= 20$ \citep{1993prpl.conf..749L,2018ApJ...861..140K}. 
Given that $M_\star / M_\bullet < 10^{-5}$, and $h/r > 10^{-2}$, $K \ll K_t$ even for quite small $\alpha$, suggesting that gap opening is not important for stars in our fiducial AGN disk. Other models have estimated that AGN disks have $h/r\sim10^{-3}-10^{-2}$ \citep[e.g.][]{2001ApJ...559..680H,2002apa..book.....F}, in which case $K\gg K_t$ is quite probable for immortal stars. In this limit, the migration timescale is increased by a factor of $\sim K/25$~\citep{2018ApJ...861..140K}, giving $\sim10^8\,\rm yr$ for $h/r=10^{-3},~M_*/M_\bullet=10^{-5},$ and $\alpha=0.1$, potentially longer than the disk lifetime.

Moreover, nonlinear and non-isothermal corotation torques \citep[e.g.][]{2010MNRAS.401.1950P, 2015ApJ...806..182D} have the potential to slow or even reverse the direction of migration for certain disk temperature and surface density gradients~\citep{Bellovary16}.
We similarly neglect heating torques, which could potentially make a large difference given that these immortal stars have near/super-Eddington luminosities~\citep{2015Natur.520...63B,2020ApJ...902...50H}.
We omit these effects because they require a more sophisticated model of the opacity and surface density in the disk than we have available, and we encourage considering heating torques in more sophisticated investigations of the stellar population.
Because we have not accounted for these effects, we are likely overestimating the magnitude of the (negative) torque on migrating stars, and significantly underestimating the migration timescale.

These uncertainties are why we did not account for migration in our population models.
This is an important aspect that we hope future work will address.

%% file: conclusion.tex
\section{Conclusions}\label{sec:conclusions}

We have estimated the population of immortal stars in AGN disks, should they exist, at $\sim 300$ for short-lived disks ($\tau \sim 0.1\rm Myr$) and $\sim 2\times 10^4$ for long-lived disks ($\tau \sim 10\rm Myr$).
In both cases these stars live in the inner $r_{\rm cap} \sim 1000-5000 r_s$, as we have only considered stellar capture and not \emph{in-situ} formation.

Taking these estimates at face value, we computed the rate of chemical pollution of AGN disks by immortal stars and found that the disk chemistry can be profoundly altered, with up to order-unity enhancements in the helium mass fraction.
This prediction, however, extrapolates far beyond the parameter space of current AGN star models, and points to the importance of exploring the evolution of stars in helium-rich AGN disks in the future.

Moreover, we expect a significant number of mergers to occur among these stars, of order $10^{-4}\mathrm{yr}^{-1}$ per AGN.
We do not know what such mergers might look like, but this seems important to investigate given ongoing and upcoming sky surveys (e.g. ZTF, LSST) which will include monitoring of at least $10^{6}\mathrm{AGN}$~\citep{2019PASP..131g8001G,2019ApJ...873..111I}.

We further predict that if stars really make up a substantial fraction of the mass of the disk then their winds may delay the dissipation of the disk, extending its lifetime by a factor of several. However, outflows from stars could drive feedback that leads to rapid dissipation of the outer disk.
It is important, therefore, to understand how these immortal stars interact with disks in a self-consistent framework, along the lines of work by~\citet{2021arXiv210707519G} for main-sequence stars.

After the AGN phase, immortal stars become mortal and should yield a population of $\sim 10M_{\odot}$ black holes \citep{2021ApJ...910...94C}. In the underlying mass function of black holes inferred by LIGO-Virgo, there is a hint of a pile-up around $10M_{\odot}$ \citep{O3a}. If this pile-up is real and associated with post-AGN stars, it implies that AGN play an important role in the black hole mergers observed with LIGO-Virgo.

Finally, we provide naive estimates of the star formation rate in AGN disks, and find that \emph{in-situ} formed stars could well outnumber captured stars by more than an order of magnitude.
More sophisticated investigations of the star formation rate therefore seem warranted, as such numbers of stars could profoundly alter the chemistry, structure and evolution of AGN disks.

While these estimates point to exciting physics and exotic processes at work in AGN disks, we emphasize their order-of-magnitude nature.
Everything we have calculated depends strongly on highly uncertain models of AGN disk structure and AGN star evolution, and the range of possibilities allowed by these uncertainties also permits e.g. chemical pollution to be quite minor, or for the population of immortal stars to be much smaller.
Moreover, the rate at which mass is used to form and grow stars is a significant fraction of the rate at which gas flows through our fiducial disk model. This highlights the need for more self-consistent models of AGN disks, which account for embedded objects. Such models may lead to substantial revision of conclusions drawn here from our order-of-magnitude estimates.

%% file: acknowledgements.tex
We are grateful to Yuri Levin for helpful comments and to Lucy Reading-Ikkanda for producing Figure~\ref{fig:schema}.
The Flatiron Institute is supported by the Simons Foundation. BM and KESF are supported by NSF AST-1831415 and Simons Foundation Grant 533845. 